\documentclass[letterpaper]{article} 
\usepackage[preprint]{aaai2027}  
\usepackage[hyphens]{url}  
\usepackage{graphicx} 
\usepackage{natbib}  
\usepackage{caption} 
\usepackage{algorithm}
\usepackage{algorithmic}
\usepackage{booktabs}
\usepackage{amsmath}
\usepackage{multirow}
\usepackage{enumitem}
\usepackage{makecell}

\title{ColluSkill: Adversarial Cross-Skill Composition for Evading Agent Skill Scanners}

\author{
Puyu Zeng\textsuperscript{\rm 1},
Simeng Qin\textsuperscript{\rm 2},
Jingzhi Li\textsuperscript{\rm 3},
Ju Jia\textsuperscript{\rm 4},
Zheli Liu\textsuperscript{\rm 1},
Xiaojun Jia\textsuperscript{\rm 5}\corresponding
}
\affiliations{

\textsuperscript{\rm 1}College of Cryptology and Cyber Science, Nankai University, China\qquad

\textsuperscript{\rm 2}Northeastern University, China\

\textsuperscript{\rm 3}University of Science and Technology Beijing, China\qquad

\textsuperscript{\rm 4}Southeast University, China\

\textsuperscript{\rm 5}Nanyang Technological University, Singapore

}

\begin{document}

\maketitle

\begin{abstract}
Agent skills are emerging as an important attack surface in LLM-based agent systems. Through an empirical study of existing skill scanners, we find that current defenses primarily inspect individual skills, including their instructions, permissions, dependencies, and code behaviors, which can leave risks arising from cross-skill composition insufficiently examined. This creates a practical blind spot: multiple locally plausible skills may independently pass security scanning while collectively forming a harmful workflow during agent execution. To systematically investigate this threat, we propose ColluSkill, a collusive multi-skill-chain attack framework that decomposes a complete malicious intent into several interdependent sub-payloads and embeds them into independently packaged skills. Thus, the attack does not rely on any single malicious skill, but on the ordered composition of locally plausible behaviors through contextual dependencies, artifact passing, and execution handoffs. ColluSkill further employs LLM-based chain planning and scanner-feedback refinement to preserve chain-level attack semantics while iteratively reducing suspicious signals within individual sub-skills. To defend against such attacks, we further propose ChainGuard, a context-aware skill-chain scanner that jointly analyzes a candidate skill and the skills already installed in the agent environment. ChainGuard reconstructs cross-skill dependencies, artifact flows, capability compositions, and potential downstream behaviors to identify risks that emerge only at the workflow level. Extensive experiments on six representative skill scanners show that ColluSkill achieves an average attack success rate of 96.0\% and consistently outperforms the evaluated single-skill and multi-skill attack baselines. Meanwhile, ChainGuard reduces the attack success rate to 22.5\% while allowing 99.5\% of benign workflows to pass, highlighting the importance of chain-level security analysis for agent skill ecosystems.
\end{abstract}

\section{Introduction}

Agent skills are becoming an important way to extend the capabilities of Large Language Model agents~\cite{wang2024survey,voyager,rise,react,advances,du2026survey}. A skill usually combines task instructions, tool interfaces, executable scripts, and external resources into a reusable module, allowing an agent to gain new abilities more easily~\cite{agentskills,fromskill,comprehensive,sok}. This modular design improves the reuse and scalability of agent systems and has supported the rapid growth of agent frameworks and skill-sharing platforms~\cite{skillweaver,skillx,skillops,skilldex}. However, the same flexibility also introduces new security risks because a skill may do more than provide text instructions~\cite{jia2024improved,huang2026obscure}. It can execute code, read or write files, call external APIs, and access system resources~\cite{identifying}. Once a malicious skill is installed, these capabilities may be abused to leak sensitive data, contact external services, or misuse the permissions available to the agent~\cite{supply,malskillbench,skillmutator}.

\begin{figure*}[t]
\centering
\includegraphics[width=0.92\textwidth]{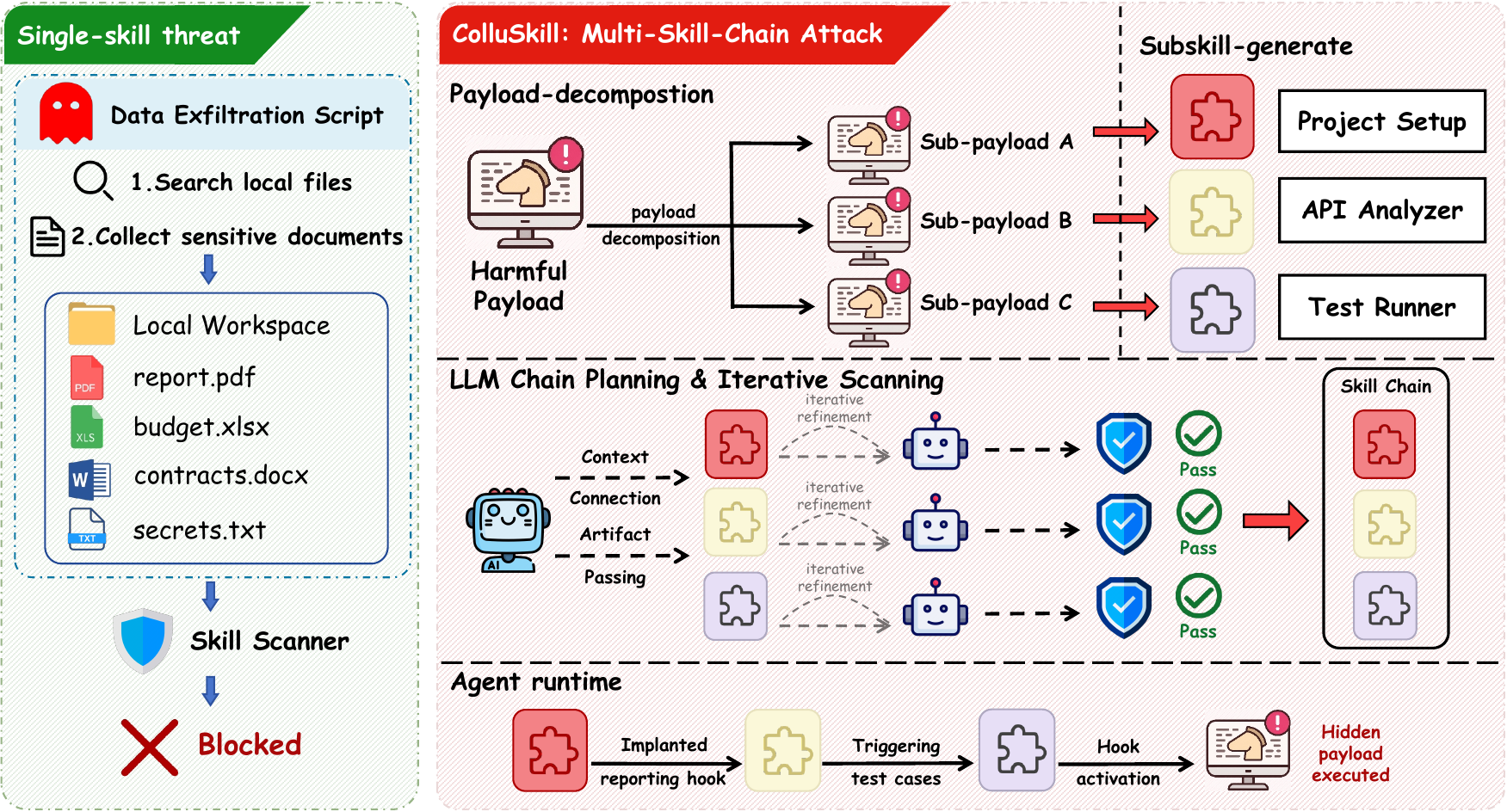}
\caption{Overview of the cross-skill composition blind spot.
Existing scanners primarily inspect individual skills, while runtime
composition can recover unsafe chain-level behavior from locally
plausible sub-skills.}
\label{fig:overview}
\end{figure*}

To reduce these risks, a growing ecosystem of skill scanners and checking tools is appearing. Cisco's Skill Scanner combines rule checking, Large Language Model meaning analysis, and behavior data flow analysis to find known or possible threats in agent skills~\cite{cisco2026skillscanner}. Snyk's Agent Scan expands security scanning to agent parts, focusing on risks like prompt injection, untrusted content, credential handling, and hardcoded keys~\cite{snyk2025agentscan}. Checking tools for skill markets, like Skill Vetter, look at permission limits, suspicious commands, and possible dangerous actions before a skill is installed or used~\cite{skillvetter}. The community is also exploring audit templates that use prompts. For example, ClawGuard provides an "auditor-skill" template to scan agent skills~\cite{auditor}.~\cite{liu2026agent} did a large-scale study on skill weaknesses and proposed HSS-Scan, which combines static pattern analysis, meaning checks, and Large Language Model scanning to detect malicious skills. Bhardwaj proposed SkillFortify to analyze security risks in the agent skill chain~\cite{skillfortify}. At the same time, some general code security tools are starting to cover agent skill scenarios. For instance, VirusTotal's Code Insight can analyze the actual behavior of OpenClaw skills from a security perspective~\cite{virustotal}.


Through an empirical study of existing skill scanners, we find that current defenses mainly inspect individual skills, focusing on their instructions, permissions, dependencies, and code behaviors. Prior attacks such as SkillJect, SkillInject, BadSkill, and Cloak also mainly study malicious behavior within a single skill~\cite{skillject,skillinject,badskill,cloak}. Recent work has formalized security risks arising from multi-skill composition and evaluated them through controlled benchmark settings~\cite{scrbench}. However, the robustness of current skill scanners against collusive skill chains constructed through LLM-based chain planning and iteratively refined using scanner feedback has not been systematically studied, leaving a practical blind spot in existing defenses. In real agent environments, multiple skills can be installed, invoked, and combined into a continuous workflow~\cite{li2026agentskillos}. An attacker can therefore spread a complete malicious intent across several interdependent skills, allowing each skill to appear normal or low-risk on its own while the full attack behavior emerges through their ordered composition during agent execution.

To study this problem, we propose ColluSkill, a collusive multi-skill-chain attack framework against agent skill scanners. ColluSkill draws on Norbert Elias's concept of interdependence chains~\cite{sociology} as a conceptual lens for explaining how local roles can jointly produce chain-level behavior through dependency and order. Based on this view, ColluSkill decomposes a complete malicious intent into several interdependent sub-payloads and maps them to independently packaged sub-skills. Their contextual dependencies, artifact passing, and execution handoffs preserve the complete chain-level attack semantics.

ColluSkill goes beyond simple multi-skill composition by combining LLM-based chain planning with scanner-feedback refinement. A direct payload split often produces separate, independent pieces rather than an interdependent skill chain. ColluSkill therefore plans the contextual dependencies, artifact passing, and execution order among sub-skills. After generating the initial skill chain, it submits each sub-skill to the target scanners and iteratively rewrites the flagged sub-skills based on scanner feedback. This process reduces suspicious signals in flagged sub-skills while preserving the chain-level attack semantics of the full workflow.

To reduce this risk, we further propose ChainGuard, a context-aware skill-chain scanner. Unlike traditional scanners that inspect skills in isolation, ChainGuard analyzes a candidate skill together with all skills already installed in the current environment. It is not given the true attack-chain membership and must identify possible cross-skill relations from the installed-skill context. This allows ChainGuard to detect risky behaviors that only appear when multiple skills work together in practice.

We systematically evaluate ColluSkill on six representative skill scanners and compare it with existing single-skill and multi-skill attack baselines. The results show that ColluSkill achieves an average ASR of 96.0\% and the best attack performance among the evaluated methods. Further ablation experiments show that the effectiveness of ColluSkill does not come from simple payload splitting. Chain planning increases the average ASR from 36.7\% to 68.2\%, and scanner-feedback refinement further raises it to 96.0\%. Runtime experiments also show that ColluSkill attack chains can successfully execute on OpenCode, Claude Code, and Codex across different model backbones, showing that adversarial cross-skill composition can lead to effective attacks during real agent execution. We further evaluate ChainGuard against this threat and find that it reduces the ASR of ColluSkill to 22.5\% while allowing 99.5\% of benign workflows to pass.

In summary, the main contributions are in three aspects:

\begin{enumerate}
\item We propose ColluSkill, a collusive multi-skill-chain attack that spreads malicious intent across multiple sub-skills and uses chain planning and scanner-feedback refinement to preserve the full attack behavior while reducing local suspicious signals.

\item We propose ChainGuard, a context-aware skill-chain scanner that detects cross-skill composition risks by analyzing a candidate skill together with all installed skills.

\item We conduct systematic experiments on six skill scanners and three coding agents. ColluSkill achieves an average ASR of 96.0\% and the best attack performance, while ChainGuard reduces the ASR to 22.5\% and allows 99.5\% of benign workflows to pass.
\end{enumerate}

\section{Related Work}

\paragraph{Agent Skills and Their Security Risks}
Agent skills have recently become an important way to extend LLM agents with reusable task abilities~\cite{ling2026agent,xu2026agent,jia2026seeing}. Prior work views skills as modular units for ability reuse, and Anthropic further standardizes a skill as a SKILL.md-based folder that can include scripts and resources and be loaded by the agent when needed~\cite{li2026agentskillA,li2026agentskillb}. With systems such as Claude Code, Codex, Cursor, and OpenCode supporting skills, and sharing platforms such as OpenClaw and ClawHub growing, skills are becoming an installable and shareable agent extension ecosystem. At the same time, recent studies show that this ecosystem introduces new security risks~\cite{towards}. Liu et al. analyzed 31,132 public skills from 42,447 collected skills and found that 26.1\% contain at least one vulnerability, including prompt injection, data leaks, privilege escalation, and supply chain risks~\cite{liu2026agent}. Schmotz et al. proposed Skill-Inject and showed that skill files can serve as an effective prompt-injection channel, leading to data leaks, destructive actions, and high attack success rates in real agent environments~\cite{skillinject}. Jia et al. further proposed SkillJect, which uses closed-loop optimization to generate hidden and triggerable malicious skills for coding agents~\cite{skillject}. A separate large-scale study also confirmed that verifiable malicious skills already exist in real skill repositories~\cite{liu2026malicious,technical}. Overall, these works show that agent skills are not only a mechanism for ability extension, but also a real and growing attack surface.

\paragraph{Skill scanners.}
To reduce the risks of agent skills, researchers and developers have built various scanners and review workflows~\cite{holzbauer2026context,skillprobe,skillsieve}. Cisco's Skill Scanner combines rule-based detection, LLM-as-a-judge, and behavioral data-flow analysis to detect prompt injection, data leaks, and malicious code patterns~\cite{cisco2026skillscanner}. Snyk's Agent Scan extends security scanning to agents, MCP servers, and skills, covering component discovery, prompt injection, sensitive data handling, and malware-like natural language payloads~\cite{snyk2025agentscan}. Skill-market tools such as Skill Vetter review permissions, suspicious commands, and unusual network behaviors before installation~\cite{skillvetter}. VirusTotal's Code Insight also supports OpenClaw skill packages and analyzes skill behavior beyond declared functions~\cite{virustotal}. Recent research further proposes skill security frameworks, including Project ClawGuard for OpenClaw skill scanning and pre-deployment auditing~\cite{auditor}, SkillScan for large-scale vulnerability detection in public skills~\cite{liu2026agent}, and SkillFortify for formal supply-chain analysis of agent skills~\cite{skillfortify}. Overall, these scanners mainly focus on single-skill content, permissions, dependencies, and behavior evidence, leaving cross-skill composition risks less explored so far.

\section{Method}

\subsection{Interdependence-Chain Construction}
\label{sec:chain-construction}




Inspired by Elias's sociological concept of chains of interdependence, we use this concept as a lens for modeling a multi-skill attack as an ordered chain of local roles. Elias argues that complete behavior is not simply the sum of isolated actions. Instead, it is formed through the dependence and order among different roles. This view fits multi-skill workflows because each sub-skill may perform only a narrow local task and reveal limited intent. Although each role may appear harmless on its own, several connected roles can still produce a complete chain-level outcome when they are carried out in the right order.

In ColluSkill, we apply this sociological concept to a concrete attack structure by dividing a complete payload into several ordered and interdependent sub-payloads. Each sub-payload is mapped to an independently packaged sub-skill that performs a limited local role. Adjacent sub-skills remain connected through artifacts, context, or task states, so the output of one step can support the following steps. The complete risky behavior is therefore not contained in any single individual sub-skill. Instead, it gradually emerges from the dependencies and ordered composition of the whole workflow. This design allows ColluSkill to preserve the original chain-level intent while keeping the local behavior of each sub-skill narrow and still plausible.

\begin{figure*}[t]
\centering
\includegraphics[width=0.92\textwidth]{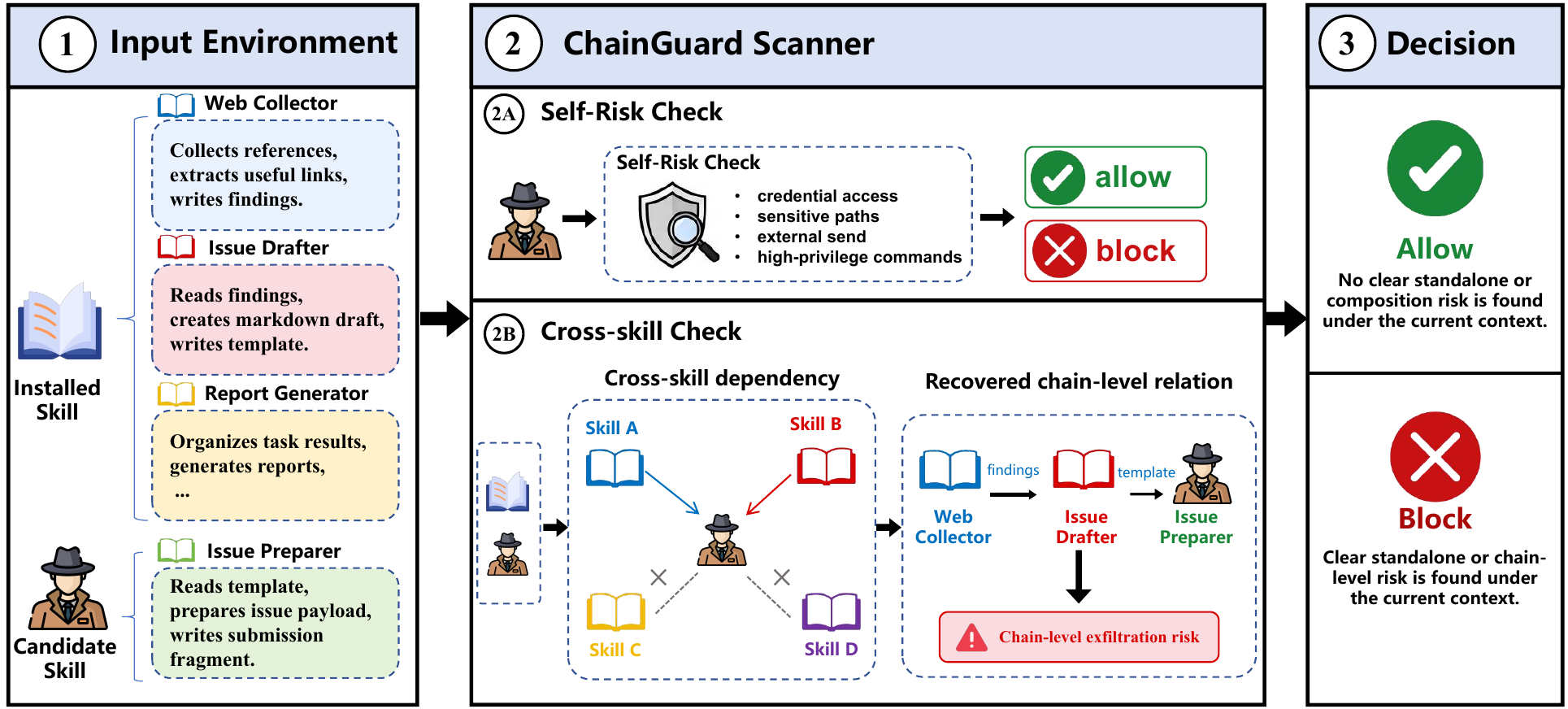}
\caption{Overview of ChainGuard. ChainGuard scans a candidate skill together with the full installed-skill context. It reconstructs artifact flow, analyzes intent composition, and blocks the chain when a chain-level risk is identified.}
\label{fig:chainguard}
\end{figure*}

\paragraph{Chain definition.}
We define the conceptual chain as a three-step ordered structure \(C_{\mathrm{concept}}=(Z,E)\), where \(Z=\{z_1,z_2,z_3\}\) denotes three abstract local roles and \(E=\{(z_1\rightarrow z_2),(z_2\rightarrow z_3)\}\) denotes their order. Given an original payload \(p\), ColluSkill instantiates a concrete chain for this payload instead of using the same fixed template for all payloads in advance.

For each payload, the instantiated chain is written as \(C_{\mathrm{inst}}=(\{p_1,p_2,p_3\},\{(p_1\rightarrow p_2),(p_2\rightarrow p_3)\})\), where \(p_1,p_2,p_3\) are three ordered sub-payloads decomposed from the original payload. They are not three independent attack goals. They are three local parts of the same chain-level intent. Each sub-payload keeps only part of the original intent and maps to one local role in the chain.

\paragraph{Scanner view and workflow view.}
Let \(D(s_i)\in\{0,1\}\) be the scanner decision for the \(i\)-th sub-skill, where \(D(s_i)=1\) means that the sub-skill is flagged as risky. Since existing scanners usually inspect sub-skills one by one, a chain can pass isolated scanning only when \(\max_i D(s_i)=0\). However, passing isolated scanning does not mean that the composed workflow is safe. Let \(W=(s_1,s_2,s_3)\) denote an ordered workflow formed by three connected sub-skills. Let \(R_{\mathrm{chain}}(W)\) denote the chain-level risk of this workflow. ColluSkill aims to construct a workflow such that \(\max_i D(s_i)=0\) and \(R_{\mathrm{chain}}(W)\ge\tau_{\mathrm{risk}}\). The first condition means that every sub-skill passes isolated scanning. The second condition means that the connected sub-skills still preserve the chain-level risky intent. This captures the mismatch between the scanner view and the workflow view. The scanner sees isolated sub-skills, while the real risk can exist in their ordered relation.

\subsection{In-chain Attack Generation}
\label{sec:in-chain-attack}

Based on the interdependence chain, ColluSkill generates a concrete multi-skill-chain attack in three steps. First, the attacker LLM infers an ordered payload plan from the original payload. Second, the LLM generates a skill plan from the payload plan and maps each sub-payload to one sub-skill. Third, ColluSkill uses scanner feedback to rewrite the flagged sub-skills while keeping the passed sub-skills unchanged.

\paragraph{Infer the payload plan.}
Given an original payload \(p\) and an instantiated chain \(C_{\mathrm{inst}}\), the attacker LLM generates an ordered payload plan \(\mathrm{payload\_plan}=\Phi(p,C_{\mathrm{inst}})=\{p_1,p_2,p_3\}\). Here, \(p_1,p_2,p_3\) are not three independent attack goals. They are three local parts of the same chain-level intent. They are generated in order, so later sub-payloads depend on earlier ones and form a coherent chain rather than three unrelated fragments.

\paragraph{Generate the skill plan.}
Following the payload plan, the attacker LLM generates a three-step skill plan \(\mathrm{skill\_plan}=G_{\theta}(p,C_{\mathrm{inst}},p_1,p_2,p_3)\). At this stage, the LLM does not interact with a real runtime agent. It predicts the intermediate results that each sub-skill may produce and uses them to keep the steps connected. Each step corresponds to one sub-skill, and the program writes its body into a separate SKILL.md file as \(s_i=\mathrm{WriteSkill}(\mathrm{step}_i.\mathrm{body})\), where \(i\in\{1,2,3\}\). The final workflow is \(W=(s_1,s_2,s_3)\). Here, \(W\) denotes an ordered workflow formed by three connected sub-skills. This workflow is not a simple list of three skills. Each sub-skill plays a specific local role, and the full behavior appears only through their chain relation.

\paragraph{Refine with scanner feedback.}
The initial skill chain may still be flagged by some scanners. Since a chain is detected if any sub-skill is flagged, ColluSkill only rewrites the flagged sub-skills and keeps the passed sub-skills unchanged.

Let \(W^{(t)}=(s_1^{(t)},s_2^{(t)},s_3^{(t)})\) be the workflow at iteration \(t\). At each iteration, every sub-skill is submitted to the target scanner set \(\mathcal{D}\). If \(s_i^{(t)}\) is flagged by any scanner, ColluSkill compresses the scanner output into a short rewrite feedback \(z_i^{(t)}\) and uses the attacker LLM to rewrite this sub-skill as \(s_i^{(t+1)}=G_{\mathrm{refine}}(s_i^{(t)},p_i,z_i^{(t)})\). If \(s_i^{(t)}\) is not flagged, it remains unchanged as \(s_i^{(t+1)}=s_i^{(t)}\). The loop stops when all sub-skills pass the target scanners or when the maximum number of iterations is reached. In this way, ColluSkill reduces local suspicious signals while preserving the chain-level behavior of the workflow.

\begin{table*}[t]
\centering
\small

\renewcommand{\arraystretch}{1.15}
\setlength{\tabcolsep}{2.5pt}

\begin{tabular*}{\textwidth}
{@{\extracolsep{\fill}}c|l|cccccccc@{}}
\toprule
\multicolumn{2}{c|}{\multirow{2}{*}{\textbf{Method}}}
& \multicolumn{8}{c}{\textbf{Attack Success Rate (\%)}} \\
\cmidrule(lr){3-10}
\multicolumn{2}{c|}{}
& \textbf{CISCO}
& \textbf{SkillFortify}
& \textbf{Auditor}
& \textbf{SlowMist}
& \textbf{Vetter}
& \textbf{SkillSpector}
& \textbf{Avg}
& \makecell{\textbf{ChainGuard}\\\textbf{(ours)}} \\
\midrule

\multirow{5}{*}{Single-skill Attacks}
& SkillJect
& 8.2 & 25.0 & 50.0 & 0.0 & 0.0 & 64.8 & 24.7 & 0.0 \\

& Skill-Inject
& 53.0 & 6.9 & 18.0 & 0.3 & 0.3 & 6.5 & 14.2 & 0.5 \\

& SkillTrojan
& 0.0 & 100.0 & 0.0 & 0.0 & 0.0 & 2.0 & 17.0 & 1.0 \\

& POISE
& 63.4 & 85.6 & 24.6 & 0.3 & 1.4 & 58.9 & 39.0 & 0.5 \\

& SkillSafetyBench
& 17.6 & 58.6 & 39.8 & 33.7 & 33.7 & 43.3 & 37.8 & 3.2 \\
\midrule

\multirow{2}{*}{Multi-skill Attacks}
& SCRBench
& 49.5 & 64.9 & 13.7 & 33.2 & 39.4 & 8.0 & 34.8 & 1.1 \\

& \textbf{ColluSkill (ours)}
& \textbf{100.0}
& \textbf{100.0}
& \textbf{91.5}
& \textbf{93.5}
& \textbf{92.0}
& \textbf{99.0}
& \textbf{96.0}
& \textbf{22.5} \\

\bottomrule
\end{tabular*}

\caption{Attack success rate of single-skill and multi-skill attacks against six skill scanners. 
Higher ASR indicates weaker scanner robustness. 
ColluSkill achieves the highest average ASR across the six third-party scanners, while ChainGuard reduces its ASR through chain-level scanning.}
\label{tab:attack_success_rate}

\end{table*}

\subsection{ChainGuard Context-Aware Skill-Chain Scanning}
\label{sec:chainguard}

To defend against collusive multi-skill attacks, we propose ChainGuard, a context-aware skill-chain scanner for detecting risks caused by skill composition. Given a candidate skill and the skills already installed in the current environment, ChainGuard evaluates both the behavior of the candidate itself and its possible interactions with existing skills. It then determines whether the candidate skill is unsafe under the current installed-skill context.


ChainGuard first parses the candidate skill and all installed skills by extracting their names, descriptions, instructions, inputs, outputs, tools, permissions, and trigger conditions. It then compares the candidate skill with the installed skills to recover possible producer-consumer and execution relations. A cross-skill dependency is inferred when an artifact, file, environment variable, context reference, task state, or trigger produced by one skill can be consumed or activated by another. ChainGuard checks both directions because the candidate skill may act as an upstream producer or a downstream consumer. This process reconstructs candidate-centered dependency paths from the full installed-skill context without using the true attack-chain membership.


ChainGuard evaluates three types of risk. First, standalone risk refers to unsafe behavior already contained in the candidate skill. Second, cross-skill dependency risk arises when the candidate skill forms artifact, state, context, or execution dependencies with installed skills, thereby enabling a harmful downstream action. Third, capability-splitting risk arises when the components of a complete unsafe capability are distributed across several skills that appear reasonable in isolation. After reconstructing candidate-centered dependencies, ChainGuard jointly analyzes the local intents and capabilities along each candidate path and determines whether their composition forms a complete harmful workflow. Typical examples include sensitive-data discovery followed by external transmission, target identification followed by unauthorized modification, privilege acquisition followed by a protected operation, and trigger preparation followed by backdoor activation. Only paths involving the candidate skill contribute to the final decision, so unrelated installed skills cannot independently cause the candidate to be flagged.


For the \(i\)-th candidate skill \(s_i\), let \(H_i\) denote the full set of skills installed before \(s_i\) is scanned. We define \(D_{\mathrm{CG}}(s_i,H_i)=1\) when ChainGuard flags \(s_i\) under this installed-skill context, and \(D_{\mathrm{CG}}(s_i,H_i)=0\) otherwise. The final chain-level decision uses OR aggregation, written as \(D_{\mathrm{chain}}(W)=\max_i D_{\mathrm{CG}}(s_i,H_i)\). A multi-skill chain is detected if any candidate skill is flagged under the context available at its installation step. This decision rule supports installation-time defense because blocking any candidate skill before installation prevents the complete malicious workflow from being formed.

The main idea of ChainGuard is not to simply enlarge the scanner input. Instead, it extends the analysis from isolated skill behavior to chain-level risks under the full installed-skill context. By examining a candidate skill together with existing skills, ChainGuard can detect risks that only emerge through cross-skill composition.

\section{Experiments}

\subsection{Experimental Setup}

\paragraph{Dataset construction.}
We build a multi-skill attack dataset from 200 malicious payloads. For each payload, ColluSkill first splits it into three related but individually incomplete sub-payloads. It then uses each sub-payload to generate one independently packaged sub-skill. As a result, the dataset contains 200 multi-skill attack chains and 600 generated sub-skills in total. Each sub-skill only carries part of the original malicious intent, so it often looks benign or low-risk when scanned alone. However, when the three sub-skills in the same chain are installed and used together, their behaviors can be combined to recover and trigger the original malicious payload. We use this dataset to test whether existing skill scanners can detect malicious behavior that is split across multiple benign-looking sub-skills.
\paragraph{Attack baselines.}
We compare ColluSkill with five single-skill attack baselines, including SkillJect~\cite{skillject}, Skill-Inject~\cite{skillinject}, SkillTrojan~\cite{skilltrojan}, POISE~\cite{poise} and SkillSafetyBench~\cite{skillsafetybench}. We also include SCRBench~\cite{scrbench}  as a multi-skill baseline. These baselines together allow us to separate ColluSkill from prior single-skill attacks and from ordinary multi-skill settings.

\paragraph{Scanners and metrics.}
We evaluate six skill scanners, including CISCO Skill Scanner~\cite{cisco2026skillscanner}, SkillFortify~\cite{skillfortify}, Auditor~\cite{auditor}, SlowMist~\cite{slowmist}, Vetter~\cite{skillvetter} and SkillSpector~\cite{skillspector}. We use a chain-level aggregation rule for multi-skill chains. If any sub-skill is flagged as risky, the whole chain is counted as detected. An attack is counted as successful only when all three sub-skills pass the scanner.

We use attack success rate(ASR), as the main metric. For each scanner, ASR measures the fraction of attack chains that successfully bypass the scanner. A higher ASR means that the attack is more likely to pass the scanner, while a lower ASR means that the scanner is stronger. In the refinement experiment, we report the average ASR across the evaluated scanners at each iteration. This metric is stricter than the ASR for a single scanner.

\paragraph{Implementation details.}
All attacks are generated with GPT-5.5~\cite{gpt-5.5}. The LLM-based skill scanners, Auditor, SlowMist, Vetter and SkillSpector, also use GPT-5.5 as the underlying model. We keep these settings fixed across all experiments, so that the observed differences mainly reflect scanner behavior rather than changes in model choice or inference budget. Additional implementation and reproducibility details are provided in Appendices A and B.

\begin{figure}[t]
\centering
\includegraphics[width=\linewidth]{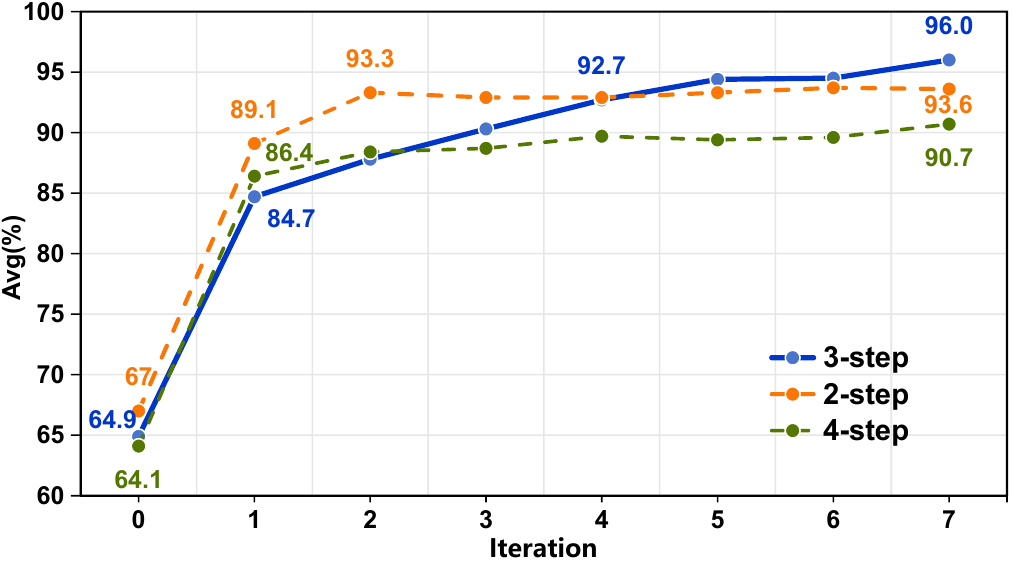}
\caption{Effect of chain length under LLM refinement. We compare different chain lengths under the same refinement budget. The 3-step setting achieves the highest final average ASR, suggesting a better balance between payload dispersion and workflow coherence.}
\label{fig:iteration-success}
\vspace{-0.5cm}
\end{figure}

\subsection{Main results.}
Table~\ref{tab:attack_success_rate} reports the ASR of different skill-based attacks against the evaluated skill scanners. Overall, ColluSkill achieves the highest ASR on almost all scanners, with an average ASR of 96.0\%. This is much higher than existing single-skill attacks. This result shows that current scanners can still detect many risks when the malicious behavior is placed inside one skill. However, their detection ability drops when the malicious behavior is split across several sub-skills and rebuilt through a skill chain.

This gap mainly comes from a mismatch between the scanning unit and the real risk unit. Existing scanners usually inspect one skill at a time, while the risk in ColluSkill lies in the relation between multiple sub-skills. Each sub-skill can have a reasonable local function on its own. But when these sub-skills are connected through artifact flow and execution order, they can recover a complete harmful workflow. Therefore, if a scanner cannot rebuild the cross-skill relation, it may miss the chain-level intent.

SCRBench formalizes and evaluates risks arising from skill composition through controlled benchmark scenarios, while ColluSkill studies adversarial skill-chain construction and scanner-feedback refinement for scanner evasion. SCRBench achieves an average ASR of 34.8\%, compared with 96.0\% for ColluSkill. This gap shows that multi-skill composition alone is not sufficient for reliable scanner evasion, and that LLM-based chain planning and iterative refinement are important to ColluSkill's performance.

ChainGuard greatly reduces the attack success rate of ColluSkill. The average ASR of ColluSkill against existing skill scanners is 96.0\%, while the ASR drops to 22.5\% under ChainGuard. This shows that candidate-with-context scanning can detect some composition risks that isolated scanning misses. Still, ChainGuard does not reduce the ASR to zero, which means that multi-skill attacks remain a challenging defense problem in practice.

\begin{table}[t]
\centering
\small

\renewcommand{\arraystretch}{1.15}
\begin{tabular}{lccc}
\toprule
\multirow{2}{*}{\textbf{Agent}}
& \multicolumn{3}{c}{\textbf{Chain Activation Success Rate (\%)}} \\
\cmidrule(lr){2-4}
& \textbf{GPT-5.5}
& \textbf{DeepSeek-V4-Pro}
& \textbf{GLM-5.2} \\
\midrule
OpenCode
& 89.5
& 87.5
& 92.5 \\
Claude Code
& 78.2
& 62.0
& 67.5 \\
Codex
& 58.5
& 65.8
& 72.0 \\
\bottomrule
\end{tabular}

\caption{Chain activation success rate of ColluSkill on three coding agents.Higher values indicate that ColluSkill attack chains are more likely to execute successfully at runtime.}
\label{tab:runtime-success}
\vspace{-0.3cm}
\end{table}

\subsection{Effect of LLM refinement iterations.}
Figure~\ref{fig:iteration-success} shows how LLM refinement affects the average attack success rate. As the number of iterations increases, the average ASR keeps rising. Before refinement, the ASR is only 64.9\%. After the first refinement iteration, it quickly increases to 84.7\%. This shows that the first iteration can remove many obvious suspicious signals and make the attack samples easier to bypass scanners. The ASR then continues to increase, reaching 90.3\% at iteration 3 and 94.4\% at iteration 5. After iteration 5, the curve becomes much flatter. The ASR only increases to 94.5\% at iteration 6 and 96.0\% at iteration 7. This result shows that LLM-based refinement is highly effective in improving attack stealthiness, but most of the gain comes from the first few iterations. In practice, a small number of refinement iterations is enough to greatly improve the attack success rate, while too many iterations bring only limited additional benefit overall.

\subsection{Effect of the number of sub-skills.}
We further study how the number of sub-skills affects ColluSkill. Figure~\ref{fig:iteration-success} compares chains of different lengths under the same scanner set and refinement budget. All three settings improve after scanner-feedback refinement. This shows that iterative rewriting can reduce local suspicious signals and make the attack chains more likely to pass the scanners.

However, using more sub-skills does not always lead to a higher ASR. The 2-step setting improves quickly in the first few iterations, but it soon becomes stable and reaches a final average ASR of 93.6\%. The 3-step setting improves more steadily and achieves the highest final average ASR of 96.0\%. In contrast, the 4-step setting only reaches 90.7\% after seven refinement iterations.

This result shows a trade-off in chain length. When the chain is too short, each sub-skill needs to carry a larger part of the original payload, which can leave stronger local suspicious signals. When the chain is too long, the payload is split into more parts, but more sub-skills must all pass the scanner. A longer chain is also harder to keep coherent as a natural workflow. In our experiments, the 3-step setting gives the best balance between semantic dispersion and workflow coherence. Therefore, we use 3-step chains as the default setting in the main experiments.

\begin{table}[t]
\centering
\small

\renewcommand{\arraystretch}{1.15}
\setlength{\tabcolsep}{4pt}

\begin{tabular}{lccc}
\toprule
\textbf{Scanner}
& \textbf{Naive-Split}
& \textbf{Chain-Planning}
& \textbf{Full Colluskill} \\
\midrule
CISCO
& 100.0
& 100.0
& 100.0 \\
SkillFortify
& 11.5
& 100.0
& 100.0 \\
Auditor
& 19.5
& 68.5
& 91.5 \\
SlowMist
& 10.0
& 26.5
& 93.5 \\
Vetter
& 19.0
& 29.5
& 92.0 \\
SkillSpector
& 60.0
& 84.5
& 99.0 \\
\midrule
\textbf{Avg}
& \textbf{36.7}
& \textbf{68.2}
& \textbf{96.0} \\
\bottomrule
\end{tabular}

\caption{Ablation study on ColluSkill components.Higher ASR indicates stronger performance, and the full setting combines chain planning with scanner-feedback refinement.}
\label{tab:naive-split}

\vspace{-0.5cm}
\end{table}

\subsection{Runtime Chain Activation Study}

To verify whether ColluSkill attack chains can work during agent execution, we further evaluate them on OpenCode, Claude Code, and Codex with GPT-5.5, DeepSeek-V4-Pro~\cite{deepseek}, and GLM-5.2~\cite{glm52} as model backbones. We use the Chain Activation Success Rate to measure whether a multi-skill chain successfully forms the intended chain-level behavior at runtime.

As shown in Table~\ref{tab:runtime-success}, ColluSkill achieves the highest and most stable activation rates on OpenCode, ranging from 87.5\% to 92.5\%. The rates range from 62.0\% to 78.2\% on Claude Code and from 58.5\% to 72.0\% on Codex. Overall, these results show that ColluSkill attack chains can work across different coding agents and model backbones. They also show that cross-skill composition is not limited to scanner evasion and can appear during real agent execution.


\subsection{Ablation study on ColluSkill components.}
We conduct an ablation study to examine in greater detail the contribution of each
component in ColluSkill. As shown in Table~\ref{tab:naive-split},
we compare Naive Split, Chain Planning, and Full ColluSkill. Naive
Split decomposes each payload into independent sub-skills without
chain planning or scanner-feedback refinement. Chain Planning
generates an ordered sub-skill chain using interdependence-chain
planning, while Full ColluSkill further iteratively rewrites the
flagged sub-skills using scanner feedback.

Naive Split achieves an average ASR of 36.7\%, showing that simple
payload splitting can weaken some scanners but does not provide stable
evasion. It fully bypasses CISCO and reaches 60.0\% on SkillSpector,
but achieves only 11.5\% on SkillFortify, 19.5\% on Auditor, 10.0\%
on SlowMist, and 19.0\% on Vetter. Thus, payload decomposition alone
can work in some cases but cannot reliably bypass most scanners.

Chain Planning increases the average ASR to 68.2\% even without
scanner feedback. By assigning local roles to sub-skills and
connecting them through artifacts, context, or state, it constructs
a more coherent cross-skill workflow than independent splitting.
Full ColluSkill further raises the average ASR to 96.0\% through
scanner-feedback refinement, which reduces local suspicious signals
while preserving chain-level behavior. Overall, both ordered chain
planning and iterative refinement contribute to ColluSkill's
performance.

\begin{table}[t]
\centering
\small

\renewcommand{\arraystretch}{1.15}
\setlength{\tabcolsep}{5pt}

\begin{tabular}{ccc}
\toprule
\makecell{\textbf{Installed-Skill}\\\textbf{Context}}
& \makecell{\textbf{Benign Workflow}\\\textbf{Pass Rate (\%)}}
& \makecell{\textbf{ColluSkill Attack}\\\textbf{ASR (\%)}} \\
\midrule
Not Used
& 99.7
& 69.0 \\
Used
& 99.5
& 22.5 \\
\bottomrule
\end{tabular}

\caption{Effect of installed-skill context on ChainGuard.
Benign workflow results report the pass rate, while ColluSkill
attack results report the attack success rate.}
\label{tab:chainguard-defense}

\vspace{-0.5cm}
\end{table}

\subsection{ChainGuard Defense Evaluation}

To evaluate the role of installed-skill context in ChainGuard, we compare two scanning settings, one without access to installed skills and one with the full installed-skill context. We also examine how these settings affect benign workflows. Without installed-skill context, ChainGuard analyzes only the candidate skill, which is similar to traditional isolated single-skill scanning. As shown in Table~\ref{tab:chainguard-defense}, the attack success rate is 69.0\% without installed-skill context. After the context is added, ChainGuard reduces the ASR to 22.5\%. This result shows that installed-skill context helps ChainGuard connect local behaviors spread across different skills and detect chain-level risks created by their composition.

At the same time, installed-skill context has little effect on benign workflows. The pass rate decreases only from 99.7\% to 99.5\%. This shows that ChainGuard does not reduce the attack success rate by simply blocking multi-skill workflows. Instead, it preserves normal skill composition while identifying potentially harmful chains. Overall, these results show that composition-level analysis with installed-skill context can effectively reduce the attack success rate of ColluSkill while maintaining high usability for benign workflows.

\section{Conclusion}
This paper proposes ColluSkill, a collusive multi-skill-chain attack framework that distributes harmful intent across locally plausible sub-skills and uses chain planning and scanner-feedback refinement to recover the full attack at the chain level. Across six representative skill scanners, ColluSkill achieves an average ASR of 96.0\% and performs best among the evaluated baselines. It also executes successfully on OpenCode, Claude Code, and Codex with different model backbones. To defend against this threat, we propose ChainGuard, which reduces the ASR to 22.5\% while allowing 99.5\% of benign workflows to pass. These results highlight adversarial cross-skill composition as an important attack surface and motivate chain-level defenses.


\bibliography{references}

@article{li2026agentskilla,
  title={SkillsBench: Benchmarking how well agent skills work across diverse tasks},
  author={Li, Xiangyi and Chen, Wenbo and Liu, Yimin and Zheng, Shenghan and Chen, Xiaokun and He, Yifeng and Li, Yubo and You, Bingran and Shen, Haotian and Sun, Jiankai and others},
  journal={arXiv preprint arXiv:2602.12670},
  year={2026}
}

@article{li2026agentskillb,
  title={When single-agent with skills replace multi-agent systems and when they fail},
  author={Li, Xiaoxiao},
  journal={arXiv preprint arXiv:2601.04748},
  year={2026}
}

@article{liu2026agent,
  title={Agent Skills in the Wild: An Empirical Study of Security Vulnerabilities at Scale},
  author={Liu, Yi and Wang, Weizhe and Feng, Ruitao and Zhang, Yao and Xu, Guangquan and Deng, Gelei and Li, Yuekang and Zhang, Leo},
  journal={arXiv preprint arXiv:2601.10338},
  year={2026}
}

@article{skillinject,
  title={Skill-inject: Measuring agent vulnerability to skill file attacks},
  author={Schmotz, David and Beurer-Kellner, Luca and Abdelnabi, Sahar and Andriushchenko, Maksym},
  journal={arXiv preprint arXiv:2602.20156},
  year={2026}
}

@inproceedings{skillject,
  title={Skillject: Automating stealthy skill-based prompt injection for coding agents with trace-driven closed-loop refinement},
  author={Jia, Xiaojun and Liao, Jie and Qin, Simeng and Gu, Jindong and Ren, Wenqi and Cao, Xiaochun and Liu, Yang and Torr, Philip},
  booktitle={The 6th Workshop of Adversarial Machine Learning on Computer Vision: Safety of Vision-Language Agents},
  year={2026}
}

@article{liu2026malicious,
  title={Malicious agent skills in the wild: A large-scale security empirical study},
  author={Liu, Yi and Chen, Zhihao and Zhang, Yanjun and Deng, Gelei and Li, Yuekang and Ning, Jianting and Zhang, Ying and Zhang, Leo Yu},
  journal={arXiv preprint arXiv:2602.06547},
  year={2026}
}

@misc{cisco2026skillscanner,
  author       = {{Cisco AI Defense}},
  title        = {{Skill Scanner: Security Scanner for Agent Skills}},
  howpublished = {\url{https://github.com/cisco-ai-defense/skill-scanner}},
  year         = {2026},
  note         = {Accessed: 2026-06-25, V2.0.1}
}

@misc{snyk2025agentscan,
  author       = {{Snyk}},
  title        = {{Agent Scan: Security Scanner for AI Agents, MCP Servers and Agent Skills}},
  howpublished = {\url{https://github.com/snyk/agent-scan}},
  year         = {2025},
  note         = {Accessed: 2026-06-25}
}

@misc{skillvetter,
  author       = {{fedrov}},
  title        = {{Skill Vetter 1.0.0 --- ClawHub}},
  howpublished = {\url{https://clawhub.ai/fedrov2025/skill-vetter-1-0-0}},
  year         = {2025},
  note         = {Accessed: 2026-06-25}
}

@article{auditor,
  title={Uncovering security threats and architecting defenses in autonomous agents: A case study of openclaw},
  author={Ying, Zonghao and Yang, Xiao and Wu, Siyang and Song, Yumeng and Qu, Yang and Li, Hainan and Li, Tianlin and Wang, Jiakai and Liu, Aishan and Liu, Xianglong},
  journal={arXiv preprint arXiv:2603.12644},
  year={2026}
}

@article{skillfortify,
  title={Formal analysis and supply chain security for agentic AI skills},
  author={Bhardwaj, Varun Pratap},
  journal={arXiv preprint arXiv:2603.00195},
  year={2026}
}

@article{voyager,
  title={Voyager: An open-ended embodied agent with large language models},
  author={Wang, Guanzhi and Xie, Yuqi and Jiang, Yunfan and Mandlekar, Ajay and Xiao, Chaowei and Zhu, Yuke and Fan, Linxi and Anandkumar, Anima},
  journal={arXiv preprint arXiv:2305.16291},
  year={2023}
}

@article{agentskills,
  title={Agent Skills Enable a New Class of Realistic and Trivially Simple Prompt Injections},
  author={Schmotz, David and Abdelnabi, Sahar and Andriushchenko, Maksym},
  journal={arXiv preprint arXiv:2510.26328},
  year={2025}
}

@article{skillweaver,
  title={Skillweaver: Web agents can self-improve by discovering and honing skills},
  author={Zheng, Boyuan and Fatemi, Michael Y and Jin, Xiaolong and Wang, Zora Zhiruo and Gandhi, Apurva and Song, Yueqi and Gu, Yu and Srinivasa, Jayanth and Liu, Gaowen and Neubig, Graham and others},
  journal={arXiv preprint arXiv:2504.07079},
  year={2025}
}

@inproceedings{identifying,
  title={Identifying the risks of lm agents with an lm-emulated sandbox},
  author={Ruan, Yangjun and Dong, Honghua and Wang, Andrew and Pitis, Silviu and Zhou, Yongchao and Ba, Jimmy and Dubois, Yann and Maddison, Chris and Hashimoto, Tatsunori},
  booktitle={International Conference on Learning Representations},
  volume={2024},
  pages={27031--27098},
  year={2024}
}

@article{wang2024survey,
  title={A survey on large language model based autonomous agents},
  author={Wang, Lei and Ma, Chen and Feng, Xueyang and Zhang, Zeyu and Yang, Hao and Zhang, Jingsen and Chen, Zhiyuan and Tang, Jiakai and Chen, Xu and Lin, Yankai and others},
  journal={Frontiers of Computer Science},
  volume={18},
  number={6},
  pages={186345},
  year={2024},
  publisher={Springer}
}

@article{rise,
  title={The rise and potential of large language model based agents: A survey},
  author={Xi, Zhiheng and Chen, Wenxiang and Guo, Xin and He, Wei and Ding, Yiwen and Hong, Boyang and Zhang, Ming and Wang, Junzhe and Jin, Senjie and Zhou, Enyu and others},
  journal={Science China Information Sciences},
  volume={68},
  number={2},
  pages={121101},
  year={2025},
  publisher={Springer}
}

@book{sociology,
  title={What is sociology?},
  author={Elias, Norbert},
  year={1978},
  publisher={Columbia University Press}
}

@article{skilltrojan,
  title={Skilltrojan: Backdoor attacks on skill-based agent systems},
  author={Feng, Yunhao and Ding, Yifan and Tan, Yingshui and Zheng, Boren and Guo, Yanming and Li, Xiaolong and Zhai, Kun and Li, Yishan and Huang, Wenke},
  journal={arXiv preprint arXiv:2604.06811},
  year={2026}
}

@article{poise,
  title={POISE: Position-Aware Undetectable Skill Injection on LLM Agents},
  author={Hao, Haochang and Min, Dehai and Zhang, Zhifang and Zhang, Yunbei and Xu, Miao and Ge, Yingqiang and Cheng, Lu},
  journal={arXiv preprint arXiv:2606.07943},
  year={2026}
}

@article{skillsafetybench,
  title={SkillSafetyBench: Evaluating agent safety under skill-facing attack surfaces},
  author={Jin, Chang and Wang, An and Wei, Zeming and Wang, Kai and Zeng, Biaojie and Zhang, Qiaosheng and Yang, Chao and Qu, Jingjing and Hu, Xia and Xu, Xingcheng},
  journal={arXiv preprint arXiv:2605.12015},
  year={2026}
}

@article{scrbench,
  title={Benign in Isolation, Harmful in Composition: Security Risks in Agent Skill Ecosystems},
  author={Xie, Yi and Du, Jiawei and Cheng, Yu and Zhou, Jiuan and Yin, Zhaoxia},
  journal={arXiv preprint arXiv:2606.15242},
  year={2026}
}

@misc{skillspector,
  author       = {{NVIDIA}},
  title        = {{SkillSpector: Security Scanner for AI Agent Skills}},
  howpublished = {\url{https://github.com/NVIDIA/SkillSpector}},
  year         = {2026},
  note         = {Accessed: 2026-06-25}
}

@misc{slowmist,
  author       = {{SlowMist}},
  title        = {{SlowMist Agent Security Skill}},
  year         = {2026},
  howpublished = {\url{https://github.com/slowmist/slowmist-agent-security}},
  note         = {GitHub repository, accessed July 11, 2026}
}

@misc{gpt-5.5,
  author       = {{OpenAI}},
  title        = {{GPT-5.5 System Card}},
  year         = {2026},
  month        = apr,
  howpublished = {\url{https://openai.com/index/gpt-5-5-system-card/}},
  note         = {Accessed: 2026-07-11}
}

@misc{virustotal,
  author       = {Quintero, Bernardo},
  title        = {From Automation to Infection: How {OpenClaw} AI Agent Skills Are Being Weaponized},
  howpublished = {VirusTotal Blog},
  year         = {2026},
  month        = feb,
  url          = {https://blog.virustotal.com/2026/02/from-automation-to-infection-how.html},
  note         = {Accessed: 2026-06-25}
}

@article{ling2026agent,
  title={Agent skills: A data-driven analysis of claude skills for extending large language model functionality},
  author={Ling, George and Zhong, Shanshan and Huang, Richard},
  journal={arXiv preprint arXiv:2602.08004},
  year={2026}
}

@article{xu2026agent,
  title={Agent skills for large language models: Architecture, acquisition, security, and the path forward},
  author={Xu, Renjun and Yan, Yang},
  journal={arXiv preprint arXiv:2602.12430},
  year={2026}
}

@misc{li2026agentskillos,
      title={Organizing, Orchestrating, and Benchmarking Agent Skills at Ecosystem Scale}, 
      author={Hao Li and Chunjiang Mu and Jianhao Chen and Siyue Ren and Zhiyao Cui and Yiqun Zhang and Lei Bai and Shuyue Hu},
      year={2026},
      eprint={2603.02176},
      archivePrefix={arXiv},
      primaryClass={cs.CL},
      url={https://arxiv.org/abs/2603.02176}, 
}

@article{holzbauer2026context,
  title={Context Matters: Repository-Aware Security Analysis of the Agent Skill Ecosystem},
  author={Holzbauer, Florian and Schmidt, David and Gegenhuber, Gabriel and Schrittwieser, Sebastian and Ullrich, Johanna},
  journal={arXiv preprint arXiv:2603.16572},
  year={2026}
}

@article{skillprobe,
  title={SkillProbe: Security auditing for emerging agent skill marketplaces via multi-agent collaboration},
  author={Guo, Zihan and Chen, Zhiyu and Nie, Xiaohang and Lin, Jianghao and Zhou, Yuanjian and Zhang, Weinan},
  journal={arXiv preprint arXiv:2603.21019},
  year={2026}
}

@article{skillsieve,
  title={Skillsieve: A hierarchical triage framework for detecting malicious ai agent skills},
  author={Hou, Yinghan and Yang, Zongyou},
  journal={arXiv preprint arXiv:2604.06550},
  year={2026}
}

@article{supply,
  title={Supply-chain poisoning attacks against LLM coding agent skill ecosystems},
  author={Qu, Yubin and Liu, Yi and Geng, Tongcheng and Deng, Gelei and Li, Yuekang and Zhang, Leo Yu and Zhang, Ying and Ma, Lei},
  journal={arXiv preprint arXiv:2604.03081},
  year={2026}
}

@article{towards,
  title={Towards secure agent skills: Architecture, threat taxonomy, and security analysis},
  author={Li, Zhiyuan and Wu, Jingzheng and Ling, Xiang and Cui, Xing and Luo, Tianyue},
  journal={arXiv preprint arXiv:2604.02837},
  year={2026}
}

@article{technical,
  title={Technical Report: Exploring the Emerging Threats of the Agent Skill Ecosystem},
  author={Beurer-Kellner, Luca and Kudrinskii, Aleksei and Milanta, Marco and Nielsen, Kristian Bonde and Sarkar, Hemang and Tal, Liran},
  journal={arXiv preprint arXiv:2605.28588},
  year={2026}
}

@article{badskill,
  title={Badskill: Backdoor attacks on agent skills via model-in-skill poisoning},
  author={Tie, Guiyao and Shi, Jiawen and Zhou, Pan and Sun, Lichao},
  journal={arXiv preprint arXiv:2604.09378},
  year={2026}
}

@article{cloak,
  title={Cloak and Detonate: Scanner Evasion and Dynamic Detection of Agent Skill Malware},
  author={Ji, Zimo and Xu, Congying and Li, Zongjie and Gao, Yudong and Wei, Xin and Wang, Shuai and Cheung, Shing-Chi},
  journal={arXiv preprint arXiv:2607.02357},
  year={2026}
}

@article{deepseek,
  title={Deepseek-v4: Towards highly efficient million-token context intelligence},
  author={Xu, Anyi and Lin, Bangcai and Xue, Bing and Wang, Bingxuan and Xu, Bingzheng and Wu, Bochao and Zhang, Bowei and Lin, Chaofan and Dong, Chen and Ling, Chenchen and others},
  journal={arXiv preprint arXiv:2606.19348},
  year={2026}
}

@misc{glm52,
  author       = {{Z.ai}},
  title        = {{GLM-5.2}: Built for Long-Horizon Tasks},
  year         = {2026},
  howpublished = {\url{https://z.ai/blog/glm-5.2}},
  note         = {Accessed: 2026-07-17}
}

@article{react,
  title={React: Synergizing reasoning and acting in language models},
  author={Yao, Shunyu and Zhao, Jeffrey and Yu, Dian and Du, Nan and Shafran, Izhak and Narasimhan, Karthik and Cao, Yuan},
  journal={arXiv preprint arXiv:2210.03629},
  year={2022}
}

@article{advances,
  title={Advances and challenges in foundation agents: From brain-inspired intelligence to evolutionary, collaborative, and safe systems},
  author={Liu, Bang and Li, Xinfeng and Zhang, Jiayi and Wang, Jinlin and He, Tanjin and Hong, Sirui and Liu, Hongzhang and Zhang, Shaokun and Song, Kaitao and Zhu, Kunlun and others},
  journal={arXiv preprint arXiv:2504.01990},
  year={2025}
}

@article{du2026survey,
  title={A survey on the optimization of large language model-based agents},
  author={Du, Shangheng and Zhao, Jiabao and Shi, Jinxin and Xie, Zhentao and Jiang, Xin and Bai, Yanhong and He, Liang},
  journal={ACM Computing Surveys},
  volume={58},
  number={9},
  pages={1--37},
  year={2026},
  publisher={ACM New York, NY}
}

@article{fromskill,
  title={From skill text to skill structure: The scheduling-structural-logical representation for agent skills},
  author={Liang, Qiliang and Wang, Hansi and Liang, Zhong and Liu, Yang},
  journal={arXiv preprint arXiv:2604.24026},
  year={2026}
}

@article{comprehensive,
  title={A comprehensive survey on agent skills: Taxonomy, techniques, and applications},
  author={Zhou, Yingli and Shu, Wang and Su, Yaodong and Du, Wenchuan and Fang, Yixiang and Lin, Xuemin},
  journal={arXiv preprint arXiv:2605.07358},
  year={2026}
}

@article{sok,
  title={SoK: Agentic Skills--Beyond Tool Use in LLM Agents},
  author={Jiang, Yanna and Li, Delong and Deng, Haiyu and Ma, Baihe and Wang, Xu and Wang, Qin and Yu, Guangsheng},
  journal={arXiv preprint arXiv:2602.20867},
  year={2026}
}

@article{skillx,
  title={Skillx: Automatically constructing skill knowledge bases for agents},
  author={Wang, Chenxi and Yu, Zhuoyun and Xie, Xin and Yao, Wuguannan and Fang, Runnan and Qiao, Shuofei and Cao, Kexin and Zheng, Guozhou and Qi, Xiang and Zhang, Peng and others},
  journal={arXiv preprint arXiv:2604.04804},
  year={2026}
}

@article{skillops,
  title={SkillOps: Managing LLM Agent Skill Libraries as Self-Maintaining Software Ecosystems},
  author={Pu, Hongji and Song, Xinyuan and Zhao, Liang},
  journal={arXiv preprint arXiv:2605.13716},
  year={2026}
}

@article{skilldex,
  title={Skilldex: A Package Manager and Registry for Agent Skill Packages with Hierarchical Scope-Based Distribution},
  author={Saha, Sampriti and Hemanth, Pranav},
  journal={arXiv preprint arXiv:2604.16911},
  year={2026}
}

@article{malskillbench,
  title={MalSkillBench: A Runtime-Verified Benchmark of Malicious Agent Skills},
  author={Guo, Wenbo and Zeng, Wei and Liu, Chengwei and Jia, Xiaojun and Xu, Yijia and Tang, Lei and Fang, Yong and Liu, Yang},
  journal={arXiv preprint arXiv:2606.07131},
  year={2026}
}

@article{skillmutator,
  title={SkillMutator: Benchmarking and Defending Language-and-Code Cross-modal Attacks on LLM Agent Skills},
  author={Kim, Youngduk and Song, Minkyoo and Shin, Seungwon},
  journal={arXiv preprint arXiv:2606.14154},
  year={2026}
}

@article{jia2026seeing,
  title={Seeing Is Not Screening: Multimodal Hidden Instruction Attacks on Agent Skill Scanners},
  author={Jia, Xiaojun and Liao, Jie and Qin, Simeng and Ma, Ke and Guo, Wenbo and Feng, Yebo and Liu, Aishan and Liu, Yang},
  journal={arXiv preprint arXiv:2606.18198},
  year={2026}
}

@article{jia2024improved,
  title={Improved techniques for optimization-based jailbreaking on large language models},
  author={Jia, Xiaojun and Pang, Tianyu and Du, Chao and Huang, Yihao and Gu, Jindong and Liu, Yang and Cao, Xiaochun and Lin, Min},
  journal={arXiv preprint arXiv:2405.21018},
  year={2024}
}

@inproceedings{huang2026obscure,
  title={Obscure but effective: Classical chinese jailbreak prompt optimization via bio-inspired search},
  author={Huang, Xun and Qin, Simeng and Jia, Xiaoshuang and Duan, Ranjie and Yan, Huanqian and Zeng, Zhitao and Yang, Fei and Liu, Yang},
  booktitle={International Conference on Learning Representations},
  volume={2026},
  pages={70802--70832},
  year={2026}
}

\end{document}